\documentstyle[12pt]{article}
\begin{document}
\hfill{CCUTH-96-06}
\vskip 0.5cm
\begin{center}
{\large {\bf The resummation approach to evolution equations}}
\vskip 1.0cm
Hsiang-nan Li
\vskip 0.5cm
Department of Physics, National Chung-Cheng University, \par
Chia-Yi, Taiwan, Republic of China
\end{center}
\vskip 2.0cm

PACS numbers: 12.38.Cy, 11.10.Hi
\vskip 4.0cm
\centerline{\bf Abstract}
\vskip 0.3cm
We derive the evolution equations of parton distribution functions 
appropriate in different kinematic regions in a unified and simple way using 
the resummation technique. They include the Gribov-Lipatov-Altarelli-Parisi 
equation for large momentum transfer $Q$, the Balitskii-Fadin-Kuraev-Lipatov 
equation for a small Bjorken variable $x$, and the 
Ciafaloni-Catani-Fiorani-Marchesini equation which unifies the above two 
equations. We propose a modified Balitskii-Fadin-Kuraev-Lipatov equation 
from the view point of the resummation. This modified version
possesses an intrinsic $Q$ dependence and its predictions for the structure 
function $F_2(x,Q^2)$ are consistent with the HERA data. 

\newpage
\centerline{\large\bf I. INTRODUCTION}
\vskip 0.5cm

Perturbative QCD (PQCD), as a gauge field theory, involves large logarithms 
from radiative corrections at each order of the coupling 
constant $\alpha_s$, such as $\ln Q$ in the kinematic region with a large 
momentum transfer $Q$ and $\ln(1/x)$ in the region with a small Bjorken 
variable $x$. These logarithms, spoiling the perturbative expansion, must be 
organized. To organize the logarithmic corrections to a parton distribution 
function, the various evolution equations have been derived. For example, 
the Gribov-Lipatov-Altarelli-Parisi (GLAP) equation \cite{AP} sums single 
logarithms $\ln Q$ for a large $x$ to all orders, the 
Balitskii-Fadin-Kuraev-Lipatov (BFKL) equation \cite{BFKL} sums $\ln(1/x)$
for a small $x$, and the Ciafaloni-Catani-Fiorani-Marchesini (CCFM) equation 
\cite{CCFM}, appropriate for both large and small $x$, unifies the above two 
equations. 

The conventional derivation of the evolution equations usually requires 
complicated diagrammatic analyses. The idea is to locate the region of 
the loop momenta flowing through the rungs (radiative gluons) of a ladder 
diagram, in which leading logarithmic corrections are produced. For the 
GLAP, BFKL, and CCFM equations, the important regions are those with the 
strong transverse momentum ordering, the strong rapidity ordering, and the 
strong angular ordering, respectively. Summing the ladder diagrams with the 
different kinematic orderings to all orders, we obtain the evolution 
equations that organize the leading logarithms. The derivation will become 
very difficult, if one intends to improve the accuracy of the evolutions to 
next-to-leading logarithms. 

In this paper we shall propose an alternative approach to the all-order 
summation of the various large logarithms. This approach is based on the 
resummation technique, which was developed originally for the organization 
of double logarithms $\ln^2 Q$ \cite{CS}. Recently, we applied this 
technique to some hard QCD processes, such as deep inelastic scattering, 
Drell-Yan production, and inclusive heavy meson decays, and demonstrated 
how to resum the double logarithms contained in parton distribution 
functions into Sudakov form factors \cite{L1}. It has been shown 
\cite{L1,L2} that the resummation technique is equivalent to the Wilson-loop 
formalism for the summation of soft logarithms. Here we shall further show 
that it can also deal with the single-logarithm cases, {\it i.e.} the 
evolution equations mentioned above. Therefore, the resummation technique 
indeed has a wide application in PQCD.

The procedures of the resummation are briefly summarized below. The 
derivative of a parton distribution function with respect to $Q$ or $x$ is 
first related to a new function, which contains a special gluon vertex. 
Expressing this new function as the convolution of the subdiagram involving 
the special vertex with the original parton distribution function, we arrive 
at the evolution equation. The subdiagram, after factorized out of the new 
function, is identified as the corresponding kernel in the different 
kinematic limits. We show explicitly that the 
lowest-order subdiagram gives the kernel for the leading-logarithm summation 
exactly. Using our approach, it is not necessary to go into the detailed 
analysis of the orderings of radiative gluons in a ladder diagrams, 
since all the possible orderings have been included in the new function.
Hence, the resummation technique provides a unified and simple 
derivation of the known evolution equations in the literature. 

In the region with both large $Q$ and small $x$ many gluons are radiated in 
scattering processes with small spatial separation among them, and a new 
effect from the annihilation of two gluons into one gluon becomes important. 
Taking into account this effect, a nonlinear evolution equation, the 
Gribov-Levin-Ryskin (GLR) equation \cite{GLR} is obtained. Using the 
resummation technique, the annihilation effect is introduced through 
next-to-leading-twist contributions to the subdiagram containing the special
vertex, and the GLR equation can also be derived easily \cite{L3}. In the 
present work we shall not address this subject, because it is not very 
relevant.

It is known that the BFKL equation is independent of $Q$, and thus its
predictions are insensitive to the variation of $Q$. However, the recent 
HERA data \cite{H1} for the structure function $F_2(x,Q^2)$ of deep 
inelastic scattering exhibit a stronger $Q$ 
dependence. It is plausible that the experiments have not yet explored the 
region with a low enough $x$ such that the BFKL equation is applicable. 
Hence, to explain the data, the $Q$ dependent GLAP equation is combined 
in some way. One can employ either the improved splitting function, which 
embeds the BFKL summation, in the GLAP equation \cite{AKMS}, or the CCFM 
equation directly \cite{KMS}, which unifies the BFKL and GLAP equations. In 
this paper we shall not rely on the GLAP summation of $\ln Q$, but propose a 
modified BFKL equation from the viewpoint of the resummation, 
which contains an intrinsic $Q$ dependence. 
It will be shown that the gluon distribution function derived from the 
modified BFKL equation gives the predictions of $F_2(x,Q^2)$, that are well 
consistent with the data.

We derive the GLAP, BFKL and CCFM equations in Sections II, III, and IV, 
respectively, by means of the resummation technique. We explain how the
subdiagram containing the special vertex reduces to the corresponding 
evolution kernels in the different kinematic regions. The modified 
BFKL equation along with its analytical solution of the gluon distribution 
function are presented in Section V. We then evaluate the structure function 
$F_2(x,Q^2)$ using the high-energy $p_T$-factorization theorem \cite{J}, and
compare them with the HERA data. Section VI is the conclusion.

\vskip 1.0cm

\centerline{\large\bf II. THE GLAP EQUATION}
\vskip 0.5cm

We derive the GLAP equation in this section. Consider deep inelastic 
scattering (DIS) of a hadron with a light-like momentum 
$p=p^+\delta^{\mu +}$ in the large $x$ limit, where 
$x=-q^2/(2p\cdot q)=Q^2/(2p\cdot q)$ is the Bjorken variable with $q$ the 
momentum transfer to the hadron through a virtual photon. It is known that 
the collinear region with loop momenta of radiative gluons parallel to 
$p$ is important, from which large logarithms arise \cite{L1}. The 
other important regions are soft, with loop momenta much smaller than 
$Q$, and hard, with loop momenta being of order $Q$. In the covariant 
gauge $\partial A=0$ the scattered quark line the collinear gluons 
attach can be replaced by an eikonal line along an arbitrary direction $n$, 
$n^2\not= 0$ \cite{L1}. The associated Feynman rules 
are $1/(n\cdot l)$ for the eikonal propagator, and $n_\mu$ for an vertex on 
the eikonal line, $l$ being the momentum flowing through it. With the 
eikonalization, the collinear gluons are factorized out of the hard 
scattering subamplitude and absorbed into a quark distribution funtion 
$\phi(\xi,p^+,\mu)$ with $\xi$ the momentum fraction, and $\mu$ a 
factorization or renormalization scale. The argument $p^+$ denotes the large 
logarithms $\ln (p^+/\mu)$ appearing in $\phi$, which will be organized 
below.

The standard definition of $\phi$ in the covariant gauge is then given by
\begin{equation}
\phi(\xi,p^+,\mu)=\int\frac{dy^-}{2\pi}e^{-i\xi p^+y^-}
\langle p| {\bar q}(y^-)\frac{1}{2}\gamma^+ Pe^{i\int_0^{y^-}dzn\cdot 
A(zn)}q(0)|p\rangle\;,
\label{dep}
\end{equation}
as shown in Fig.~1(a), where $\gamma^+$ is a Dirac matrix, and $|p\rangle$ 
denotes the incoming hadron. The path-ordered exponential 
$Pe^{i\int dzn\cdot A}$ represents exactly the eikonal line collecting the 
collinear gluons. It is easy to confirm that this exponential 
generates the Feynman rules stated above. Though $\phi$ constructed 
in this way contains an artificial dependence on $n$, the cross 
section of the scattering process, being a physical quantity, should be
$n$ independent. It has been shown that the $n$ dependence of $\phi$ is 
canceled by those of other involved subprocesses \cite{L1}. This vector $n$ 
is, however, essential at the intermediate stage of the resummation. 

In the axial gauge $n\cdot A=0$ the path-ordered exponential is equal to
the identity, and $\phi$ is defined by Fig.~1(b), implying that the 
collinear gluons are decoupled from the eikonal line. The $n$ dependence 
then goes into the gluon propagator, $(-i/l^2)N^{\mu\nu}(l)$, with
\begin{equation}
N^{\mu\nu}=g^{\mu\nu}-\frac{n^\mu l^\nu+n^\nu l^\mu}
{n\cdot l}+n^2\frac{l^\mu l^\nu}{(n\cdot l)^2}\;.
\label{gp}
\end{equation}
The above decoupling can also be understood from the relation 
$n_\mu N^{\mu\nu}=0$ associated with the attachment of a radiative gluon to 
the eikonal line. We have shown that the resummation results derived in the 
axial and covariant gauges are the same \cite{L1}, and thus we can work in 
either of them. In this section we adopt the axial gauge.

The key step of the resummation is to obtain the derivative $p^+d\phi/dp^+$. 
Because of the scale invariance of $\phi$ in the vector $n$ as indicated by 
Eq.~(\ref{dep}) or (\ref{gp}), $\phi$ must depend on $p$ and $n$ through the 
ratio $(p\cdot n)^2/n^2$. Hence, we have the chain rule relating $p^+d/dp^+$ 
to $d/dn$: 
\begin{eqnarray}
p^+\frac{d}{dp^+}\phi=-\frac{n^2}{v\cdot n}v_\alpha\frac{d}{dn_\alpha}
\phi\;,
\label{cph}
\end{eqnarray}
with $v_\alpha=\delta_{\alpha +}$ a vector along $p$. In the axial gauge 
$d/dn$ applies only to the gluon propagator, giving
\begin{equation}
\frac{d}{dn_\alpha}N^{\mu\nu}=
-\frac{1}{n\cdot l}(l^\mu N^{\alpha\nu}+l^\nu N^{\mu\alpha})\;.
\label{dgp}
\end{equation}
The loop momentum $l^\mu$ ($l^\nu$) flowing through the differentiated gluon 
line contracts with the vertex the gluon attaches, which is then 
replaced by a special vertex 
\begin{eqnarray}
{\hat v}_\alpha=\frac{n^2v_\alpha}{v\cdot nn\cdot l}\;.
\label{va}
\end{eqnarray}
This special vertex can be simply read off the combination of 
Eqs.~(\ref{cph}) and (\ref{dgp}).

The contraction of $l^\mu$ ($l^\nu$) hints the application of the Ward 
identity, from which it is found that the special vertex moves to the outer 
end of the valence quark line. For a given $x$, we obtain the 
formula, 
\begin{equation}
p^+\frac{d}{dp^+}\phi(x,p^+,\mu)=2{\tilde \phi}(x,p^+,\mu)\;,
\label{dph}
\end{equation}
shown in Fig.~2(a), where $\tilde \phi$, the new function mentioned in the
Introduction, contains one special vertex represented by a square. The 
coefficient 2 comes from the equality of the new functions with the special
vertex on either of the two valence quark lines. Note that Eq.~(\ref{dph}) 
is an exact consequence of the Ward identity without specifying the ordering 
of radiative gluons. All the possible orderings are embedded in 
Eq.~(\ref{dph}).

We show that Eq.~(\ref{dph}) leads to the GLAP equation, when $p^+$, or
equivalently $Q$ in the center-of-mass frame of the virtual photon and the
incoming quark, is large. The important regions of the loop momentum 
flowing through the special vertex are soft and hard, since the vector $n$ 
does not lie on the light cone, and the collinear enhancements are 
suppressed. In the important soft and hard regions $\tilde \phi$ can be 
factorized into the convolution of the subdiagram containing the special 
vertex with the original distribution function $\phi$,
\begin{eqnarray}
{\tilde \phi}(x,p^+,\mu)=
\int_x^1 d\xi [K(x,\xi,p^+,\mu)+G(x,\xi,p^+,\mu)]\phi(\xi,p^+,\mu)\;.
\label{apeq}
\end{eqnarray}
The function $K$, absorbing the soft divergences of the subdiagram, 
corresponds to Fig.~2(b), 
where the eikonal approximation for the valence quark propagator has been 
made. The function $G$, absorbing the ultraviolet divergences, 
corresponds to Fig.~2(c), where the subtraction of the second diagram
ensures that the involved loop momentum is of order $Q$. 

We identify the factor $K+G$ as the GLAP kernel. According to 
Figs.~2(b) and 2(c), $K$ and $G$ are written as
\begin{eqnarray}
K&=&ig^2{\cal C}_F\mu^\epsilon\int
\frac{d^{4-\epsilon}l}{(2\pi)^{4-\epsilon}}
\frac{{\hat v}_\mu v_\nu}{v\cdot l}
\left[\frac{\delta(\xi-x)}{l^2}+2\pi i\delta(l^2)\delta(\xi-x-l^+/p^+)
\right]N^{\mu\nu}
\nonumber \\
& &-\delta K\;,
\label{kj}\\
G&=&-ig^2{\cal C}_F\mu^\epsilon\int\frac{d^{4-\epsilon}l}
{(2\pi)^{4-\epsilon}}{\hat v}_\mu
\left[\frac{\xi\not p-\not l}{(\xi p-l)^2}\gamma_\nu
+\frac{v_\nu}{v\cdot l}\right]\frac{N^{\mu\nu}}{l^2}\delta(\xi-x)
-\delta G\;,
\label{gph}
\end{eqnarray}
where ${\cal C}_F=4/3$ is the color factor. The $\delta$ functions come 
from the final state cut, and $\delta K$ and $\delta G$ are additive 
counterterms. The virtual and real gluon emissions correspond to the first 
and second terms in the integral of $K$, respectively. The second term in 
the integral of $G$ is the soft subtraction. We emphasize that the 
factorization formula in Eq.~(\ref{apeq}) is not an exact relation, but 
holds only up to the leading logarithms $\ln p^+$. As explained later, the
approximation equivalent to the strong transverse momentum ordering in the 
conventional approach has been applied to Eq.~(\ref{kj}).

A straightforward calculation gives
\begin{eqnarray}
K&=&\frac{\alpha_s(\mu)}{\pi\xi}{\cal C}_F\left[\frac{1}{(1-x/\xi)_+}
+\ln\frac{\nu p^+}{\mu}\delta(1-x/\xi)\right]\;,
\nonumber\\
G&=&-\frac{\alpha_s(\mu)}{\pi\xi}{\cal C}_F\ln\frac{\xi\nu p^+}{\mu}
\delta(1-x/\xi)\;,
\label{kgir}
\end{eqnarray}
where constants of order unity have been dropped, and
$\nu=\sqrt{(v\cdot n)^2/|n^2|}$ is the gauge factor. Equation (\ref{kgir})
confirms our argument that $\phi$ depends on $p$ and $n$ through the ratio
$(p\cdot n)^2/n^2=(\nu p^+)^2$. In the considered region with $x\to 1$ 
the logarithm $\ln(\xi\nu p^+/\mu)$ in $G$ can be replaced by
$\ln(\nu p^+/\mu)$.

We then treat $K$ and $G$ by renormalization group (RG) methods:
\begin{equation}
\mu\frac{d}{d\mu}K=-\lambda_K=-\mu\frac{d}{d\mu}G\;.
\label{kg}
\end{equation}
The anomalous dimension of $K$ is defined by $\lambda_K=-\mu d\delta K
/d\mu$, whose explicit expression is not essential here. When solving 
Eq.~(\ref{kg}), we allow the variable $\mu$ to evolve from the scale of $K$ 
to the scale of $G$. The RG solution of $K+G$ is given by
\begin{eqnarray}
K(x,\xi,p^+,\mu)+G(x,\xi,p^+,\mu)&=&K(x,\xi,p^+,p^+)+G(x,\xi,p^+,p^+)
\nonumber \\
& &-\int_{p^+}^{p^+}\frac{d{\bar\mu}}{\bar\mu}
\lambda_K(\alpha_s({\bar\mu}))\;,
\nonumber \\
&=&\frac{\alpha_s(p^+)}{\pi\xi}{\cal C}_F\frac{1}{(1-x/\xi)_+}\;.
\label{skg}
\end{eqnarray}
It is obvious that the source of double logarithms, {\it i.e.} the integral 
cantaining $\lambda_K$, vanishes. 

A remark is in order. The function $\delta(\xi-x-l^+/p^+)$ in 
Eq.~(\ref{kj}) will be replaced by $\delta(\xi-x)\exp(i{\bf l}_T\cdot 
{\bf b})$, $b$ being the conjugate variable of the transverse momentum 
carried by the valence quark, if the transverse degrees of freedom are 
considered \cite{L1}. The integration over $\xi$ can thus be performed 
trivially, and the convolution form in Eq.~(\ref{apeq}) is simplified to a 
multiplication form,
\begin{eqnarray}
{\tilde \phi}(x,b,p^+,\mu)=[K(x,b,\mu)+G(x,p^+,\mu)]\phi(x,b,p^+,\mu)\;.
\label{apm}
\end{eqnarray}
This simplication will be implemented in Sec. V to solve the modified
BFKL equation. In this case the scale $1/b$, instead of $p^+$, serves as the 
infrared cutoff of $K$. Consequently, $1/b$ is substituted for the lower 
bound of $\bar\mu$ in Eq.~(\ref{skg}). Double logarithms then exist, 
implying that the soft logarithms in $\phi$ do not cancel completely. 
Therefore, the resummation technique can deal with both the single-logarithm 
and double-logarithm problems.

Inserting Eq.~(\ref{skg}) into (\ref{apeq}) and solving (\ref{dph}), 
we obtain 
\begin{equation}
\phi(x,\mu)=\phi(x,\Lambda,\mu)+\int_\Lambda^\mu \frac{d{\bar\mu}}{\bar\mu}
\frac{\alpha_s({\bar\mu})}{\pi}{\cal C}_F
\int_x^1 \frac{d\xi}{\xi} \frac{2}{(1-x/\xi)_+}\phi(\xi,{\bar\mu},\mu)\;,
\label{sp1}
\end{equation}
with $\Lambda$ an arbitrary cutoff. We have defined 
$\phi(x,\mu)\equiv\phi(x,\mu,\mu)$, which does not contain large logarithms.
According to the formalism in \cite{L1}, the upper bound of $\bar\mu$ 
should be set to the hadron momentum. However, it is equivalent to 
set it to $\mu$ here. It is known that a physical quantity such as the 
DIS cross section is $\mu$ independent. When 
applying the RG analysis to the factorization formula for the cross section, 
which is a convolution of the hard scattering subamplitude with $\phi$, 
$\mu$ evolves to the characteristic scale of the hard part at last. This 
hard scale is exactly the large hadron momentum. 

Differentiating Eq.~(\ref{sp1}) with respect to $\mu$, and substituting 
the RG equation $\mu d\phi(x(\xi),\Lambda({\bar \mu}),\mu)/d\mu=-2\lambda_q
\phi(x(\xi),\Lambda({\bar \mu}),\mu)$, $\lambda_q=-\alpha_s/\pi$ 
being the quark anomalous dimension in the axial gauge, we have
\begin{eqnarray}
\mu\frac{d}{d\mu}\phi(x,\mu)=
\frac{\alpha_s(\mu)}{\pi}{\cal C}_F\int_x^1 \frac{d\xi}{\xi} 
\frac{2}{(1-x/\xi)_+}\phi(\xi,\mu)
-\lambda_q(\mu)\phi(x,\mu)\;.
\end{eqnarray}
The above equation can be reexpressed as
\begin{eqnarray}
Q\frac{d}{dQ}\phi(x,Q)=\frac{\alpha_s(Q)}{\pi}
\int_x^1 \frac{d\xi}{\xi} P(x/\xi)\phi(\xi,Q)\;,
\label{sp2}
\end{eqnarray}
where $\mu$ has been set to $Q$ and the kernel $P$ is 
\begin{eqnarray}
P(x)={\cal C}_F\left[\frac{2}{(1-x)_+}+\frac{3}{2}\delta(1-x)\right]\;.
\end{eqnarray}
It is trivial to identify $P$ as the splitting function $P_{qq}$ in the 
limit $x\to 1$,
\begin{eqnarray}
P_{qq}(x)={\cal C}_F\left[\frac{1+x^2}{(1-x)_+}
+\frac{3}{2}\delta(1-x)\right]\;.
\end{eqnarray}
Hence, Eq.~(\ref{dph}) leads to the GLAP equation (\ref{sp2}). 

In fact, only the terms of $P_{qq}$, which are singular at $x\to 1$,
can be reproduced in the resummation formalism. We emphasize 
that the factorization of the subdiagram containing the special vertex 
makes sense only in the leading soft and hard regions, and thus the 
functions $K$ and $G$ collect only the most important contributions to the 
splitting function. This conclusion applies to other cases that involve
a soft approximation, such as the CCFM equation in the conventional 
derivation \cite{CCFM}, where the finite part of the relevant splitting 
function was also missing and put in by hand at last.

\vskip 1.0cm

\centerline{\large\bf III. THE BFKL EQUATION}
\vskip 0.5cm

In this section we demonstrate that the resummation technique reduces to the 
BFKL equation for the gluon distribution function in the small $x$ region. 
It is convenient to adopt the covariant gauge $\partial A=0$, under which 
the unintegrated gluon distribution function $F(x,p_T)$ is defined by 
Fig.~1(a) with the valence partons being gluons. The path-ordered 
exponential in Eq.~(\ref{dep}), and thus the eikonal line in the direction 
$n$, which collects radiative gluons, appear. $F(x,p_T)$ describes the 
probability of a gluon carrying a longitudinal momentum fraction $x$ and a 
transverse momentum $p_T$. We have made explicit that $F$ depends on 
$p_T$, instead of the large hadron momentum $p^+$ appearing in the GLAP case. 
It will be shown later that $p^+$ disappears as $x\to 0$, and thus the $p_T$ 
dependence is not negligible. For a unified treatment of the large-$x$ and 
small-$x$ summation, the $p^+$ dependence should be included, and the 
resummation technique reduces to the CCFM equation discussed in Sec. IV. 

To sum large logarithms $\ln(1/x)$, the strong rapidity ordering of 
radiative gluons in a ladder diagram is assumed in the conventional 
approach. In the resummation formalism the BFKL equation can be 
derived simply by reinterpreting the derivative with respect to $p^+$ in 
Eq.~(\ref{dph}) and by modifying the expression of $K$ in Eq.~(\ref{kj}). 
Though $F$ does not depend on $p^+$ explicitly, it can vary 
with $p^+$ through the momentum fraction implicitly, which is proportional 
to $(p^+)^{-1}$ for a fixed parton momentum. For a similar reason, $F$ 
depends on the ratio $(p\cdot n)^2/n^2$, and thus Eq.~(\ref{cph}) holds. In 
the covariant gauge the operator $d/dn$ applies to the Feynman rules for the 
eikonal line, giving
\begin{equation}
\frac{d}{dn_\alpha}\frac{n^\mu}{n\cdot l}=\frac{1}{n\cdot l}
\left(g^{\mu\alpha}-\frac{n^\mu l^\alpha}{n\cdot l}\right)\;.
\label{del}
\end{equation}
Combining Eqs.~(\ref{cph}) and (\ref{del}), we find that the differentiation 
with respect to $p^+$ generates a special vertex on the eikonal line,
\begin{eqnarray}
{\hat n}_\alpha=\frac{n^2}{v\cdot n}\left(\frac{v\cdot l}{n\cdot l}n_\alpha
-v_\alpha\right)\;.
\label{dp}
\end{eqnarray}

The derivative of $F$ is then expressed as 
\begin{equation}
p^+\frac{d}{dp^+}F(x,p_T)\equiv -x\frac{d}{dx}F(x,p_T)
=4{\tilde F}(x,p_T)\;,
\label{df}
\end{equation}
described by Fig.~3(a), where the new function $\tilde F$ contains one
special vertex denoted by the symbol $\times$. Note that the coefficient 4 
in front of $\tilde F$ is twice of the corresponding coefficient in the GLAP 
case. Since the gluon interacts with the virtual photon 
through a quark box, two quark lines, and thus two eikonal lines after 
factorizing out the gluon distribution function, are adjacent 
to the parton vertex. Hence, there is one more attachment of the special 
vertex to the eikonal line on each side of the final state cut. It is 
trivial to show that these attachments 
give the same results. Starting with Eq.~(\ref{df}), we need not to go into 
the detailed analysis of the rapidity ordering of radiative gluons,
since all the possible orderings have resided in it. 

The leading regions of the loop momentum $l$ flowing through the special 
vertex are also soft and hard. If $l$ is collinear, the first term 
$v\cdot l$ in ${\hat n}_\alpha$ vanishes, and the second term $v_\alpha$, 
as contracted with a vertex in the distribution function which is dominated 
by momenta parallel to $p$, gives a small contribution. The soft divergences 
of the subdiagram are collected by Fig.~3(b), 
and the ultraviolet divergences by Fig.~3(c). We employ the relation 
$f_{abc}t_bt_c=(i/2)Nt_a$ for the color structure, $t$ being the color 
matrices and $N=3$ being the number of colors. Absorbing $t_a$ into 
the parton vertex, Fig.~3(b) leads to                   
\begin{eqnarray}
{\tilde F}_{\rm soft}(x,p_T)&=&
-\frac{i}{2}Ng^2\int\frac{d^{4}l}{(2\pi)^4}
\frac{\Gamma^{\mu\nu\lambda}{\hat n}_\nu}{(2v\cdot l)n\cdot l}
\left[2\pi i\delta(l^2)F(x,|{\bf p}_T+{\bf l}_T|)\right.
\nonumber \\
& &\left.+\frac{\theta(p_T^2-l_T^2)}{l^2}F(x,p_T)\right]\;,
\label{kf}
\end{eqnarray}
where the triple-gluon vertex for vanishing $l$ is given by
\begin{equation}
\Gamma^{\mu\nu\lambda}=
-g^{\mu\nu}v^{\lambda}-g^{\nu\lambda}v^{\mu}+2g^{\lambda\mu}v^{\nu}\;.
\label{tri}
\end{equation}
The first term in the above integral corresponds to the real gluon emission, 
where $F(x,|{\bf p}_T+{\bf l}_T|)$ indicates that the parton coming out of
the hadron carries a transverse momentum ${\bf p}_T+{\bf l}_T$ in order to 
radiate a real gluon of momentum ${\bf l}_T$. The second term corresponds to 
the virtual gluon emission, where the $\theta$ function sets the upper bound 
of $l_T$ to $p_T$ to ensure a soft momentum flow. Therefore, it is not 
necessary to introduce a renormalization scale $\mu$ into $F$. There is not
the $\xi$ integration in Eq.~(\ref{kf}), since we have included the 
transverse degrees of freedom of the parton as a soft regulator, and the 
final state cut associated with the real gluon can be approximated by 
$\delta(\xi-x-l^+/p^+)\approx \delta(\xi-x)$ as stated in the previous 
section. 

It can be easily shown that $v^\lambda$ in Eq.~(\ref{tri}), contracted with 
a vertex in the quark box diagram, leads to a contribution
smaller by a power $1/s$ with $s=(p+q)^2$, compared to the contribution 
from the last term $v^\nu$. Similarly, the term $v^\mu$ is also contracted
with a vertex in the box diagram through the metric tensor 
associated with the gluon distribution function. Hence, we drop the first 
two terms of $\Gamma^{\mu\nu\lambda}$, and absorb $g^{\lambda\mu}$ into
$F$. Evaluating the integral straightforwardly, Eq.~(\ref{kf}) reduces to              
\begin{eqnarray}
{\tilde F}_{\rm soft}(x,p_T)=
\frac{{\bar \alpha}_s}{4}
\int\frac{d^{2}l_T}{\pi l_T^2}
\left[F(x,|{\bf p}_T+{\bf l}_T|)-\theta(p_T^2-l_T^2)F(x,p_T)\right]\;,
\label{kf1}
\end{eqnarray}
with ${\bar \alpha}_s=N\alpha_s/\pi$.

It was argued that when the fractional momentum of a parton vanishes, 
the associated collinear enhancements are suppressed \cite{L1}. The 
vanishing of the contribution from the first diagram of Fig.~3(c)
\begin{eqnarray}
G^{(1)}&=&\frac{{\bar\alpha}_s}{4}\int\frac{d^{2}l_T}{\pi}
\left[\frac{1}{l_T^2}-\frac{1}{l_T^2+(xp^+\nu)^2}\right.
\nonumber \\
& &\left.
-\frac{1}{2}\frac{xp^+\nu}{[l_T^2+(xp^+\nu)^2]^{3/2}}
\ln\frac{\sqrt{l_T^2+(xp^+\nu)^2}-xp^+\nu}
{\sqrt{l_T^2+(xp^+\nu)^2}+xp^+\nu}\right]
\end{eqnarray}
at $x\to 0$, reflects this argument. Hence, $F$ does not acquire a 
dependence on the large scale $p^+$, and the transverse degrees of freedom 
must be taken into account, differing from the GLAP case for a large $x$. 
This is the basic idea of the so-called high-energy $p_T$-factorization 
theorem \cite{J}. Therefore, the introduction of $p_T$ and the 
disappearence of $p^+$ are built in the resummation technique naturally. It 
is obvious that our formalism is applicable to the distribution functions 
constructed according to the collinear factorization (the GLAP case) and 
according to the $p_T$-factorization (the BFKL case). 

Neglecting $G^{(1)}$ along with its soft subtraction (the second diagram
in Fig.~3(c)), that is, adopting ${\tilde F}={\tilde F}_{\rm soft}$, 
Eq.~(\ref{df}) becomes 
\begin{eqnarray}
\frac{F(x,p_T)}{d\ln(1/x)}=
{\bar \alpha}_s\int\frac{d^{2}l_T}{\pi l_T^2}
\left[F(x,|{\bf p}_T+{\bf l}_T|)-\theta(p_T^2-l_T^2)F(x,p_T)\right]\;,
\label{bfkl}
\end{eqnarray}
which is exactly the BFKL equation. It is then understood that the 
subdiagram containing the special vertex plays the role of the BFKL kernel.

In summary, the BFKL equation is appropriate for the multi-Regge region, 
where the transverse momenta flowing through the rungs of a ladder diagram 
are of the same order, {\it i.e.} $l_T\approx p_T$. Hence, the loop
momentum $l_T$ flowing through the parton distribution function is not
negligible, and the final state cut $\delta(\xi-x-l^+/p^+)$ can be 
approximated by $\delta(\xi-x)$ as in Eq.~(\ref{kf}). While the 
GLAP equation is appropriate for the transverse momemtum ordered region, in 
which we have $l_T\ll p_T$, {\it i.e.} $F(x,|{\bf p}_T+{\bf l}_T|)\approx
F(x,p_T)$ for the real gluon emission. The $p_T$ dependence of the parton 
distribution function then
decouples, and can be integrated out from both sides of Eq.~(\ref{bfkl}). 
Hence, a parton distribution function in the GLAP equation does not 
involve the transverse degrees of freedom. In this case the loop momentum 
$l^+$ should be maintained in the final state cut as in 
Eq.~(\ref{kj}), which then leads to the splitting function. If not, the 
right-hand side of Eq.~(\ref{kj}) will be identical to zero. Therefore,
the subdiagram containing the special vertex reduces to the corresponding 
evolution kernels in the different kinematic regions.

\vskip 1.0cm

\centerline{\large\bf IV. THE CCFM EQUATION}
\vskip 0.5cm

Based on the discussion in the previous two sections, it is not difficult
to demonstrate that Eq.~(\ref{df}) reduces to the CCFM equation \cite{CCFM}, 
which is appropriate for both large $x$ and small $x$. It embodies the 
GLAP equation and the BFKL equation, and depends on the longitudinal 
momentum $p^+$ and the transverse momentum $p_T$ of a parton at the same 
time. For the conventional derivation of the CCFM equation, assuming the 
angular ordering of radiative gluons in a ladder diagram, refer to 
\cite{CCFM}. By means of the resummation technique, the complicated 
diagrammatic analysis can be avoided, and the physical meaning of each 
factor in the CCFM equation is clearer. 

Consider Eq.~(\ref{df}) but with the unintegrated gluon distribution 
function $F$ depending on $p_T$ and $p^+$,
\begin{equation}
p^+\frac{d}{dp^+}F(x,p_T,p^+)=4{\tilde F}(x,p_T,p^+)\;,
\label{cc}
\end{equation}
which manifests the attempt to unify the GLAP and BFKL equations. It is not
necessary to introduce a renormalization scale $\mu$ here, because the 
transverse degrees of freedom will not be integrated out. Again, the new 
function $\tilde F$ involves one special vertex on the eikonal line in the 
direction $n$.

If following the standard procedures of the resummation, we should factorize 
the subdiagram containing the special vertex by absorbing its soft and hard 
contributions into the functions $K$ and $G$, respectively. The $p^+$ 
dependence of $F$ then comes from $G$, which collects the virtual gluon 
emissions. This idea leads to a new unified evolution equation, which will 
be studied elsewhere. To reproduce the CCFM equation, however, the inclusion 
of the virtual corrections must be performed in a different way: We organize 
the virtual gluons embedded in $K$, instead of those in $G$. 
Hence, the subdiagram is factorized into Fig.~4(a), where the two jet 
functions $J$ group all the possible virtual corrections, and the real gluon
between them is soft. It will be shown below that this subdiagram gives
the CCFM kernel.

First, we resum the double logarithms contained in $J$ by considering its
derivative
\begin{eqnarray}
p^+\frac{d}{dp^+}J(p_T,p^+)&=&2{\tilde J}(p_T,p^+)
\nonumber \\
&=&2[K_J(p_T,\mu)+G_J(p^+,\mu)]J(p_T,p^+)\;.
\label{ccj}
\end{eqnarray}
At lowest order the function $K_J$ comes from the first diagram of 
Fig.~3(b), and $G_J$ from Fig.~3(c). The coefficient 2 counts the two
eikonal lines adjacent to the parton vertex. The relation between $K_J+G_J$ 
and $J$ is simply multiplicative, since $J$ groups only virtual gluons. 
We have set the infrared cutoff of $K_J$ to $p_T$, as indicated by its 
argument. This cutoff is necessary here due to the lack of the corresponding 
real gluon emission, which serves as a soft regulator. The one-loop $K_J$ 
can be easily obtained by working out the second integral in 
Eq.~(\ref{kf1}) without the $\theta$ function. The anomalous dimension of 
$K_J$ is then found to be $\gamma_J={\bar\alpha_s}/2$. The function $G_J$ 
can also be computed, but its explicit expression is not important. The 
standard RG analysis gives
\begin{equation}
K_J(p_T,\mu)+G_J(p^+,\mu)=
-\int_{p_T}^{p^+}\frac{d{\bar\mu}}{\bar\mu}\gamma_J(\alpha_s({\bar\mu}))\;,
\label{cckg}
\end{equation}
with the initial conditions $K_J(p_T,p_T)=G_J(p^+,p^+)=0$. Of course, we 
have neglected the constants of order unity in $K_J$ and $G_J$.

Substituting Eq.~(\ref{cckg}) into (\ref{ccj}), we solve for 
\begin{equation}
J(p_T,Q)=\Delta(Q,p_T)J^{(0)}\;,
\end{equation}
with the double-logarithm exponential 
\begin{eqnarray}
\Delta(Q,p_T)=\exp\left[-{\bar\alpha_s}
\int_{p_T}^{Q}\frac{dp^+}{p^+}
\int_{p_T}^{p^+}\frac{d{\bar\mu}}{\bar\mu}\right]\;.
\end{eqnarray}
We have chosen the upper bound of $p^+$ as $Q$, and 
ignored the running of ${\bar\alpha}_s$. The initial condition $J^{(0)}$, 
regarded as the tree-level gluon propagator, will appear in the integrand 
for the real gluon emission in Fig.~4(b). We split the above exponential 
into 
\begin{equation}
\Delta(Q,p_T)=
\Delta_S^{1/2}(Q,zq)\Delta_{NS}^{1/2}(z,q,p_T)\;, 
\end{equation}
with $z=x/\xi$ and $q=l_T/(1-z)$, where $\xi$ is the momentum fraction
entering $J$ from the bottom, and $l_T$ is the transverse loop momentum 
carried by the real gluon. The so-called ``Sudakov" exponential
$\Delta_S$ and the ``non-Sudakov" exponentials $\Delta$ are given by
\begin{eqnarray}
\Delta_S(Q,zq)&=&\exp\left[-2{\bar\alpha_s}
\int_{zq}^{Q}\frac{dp^+}{p^+}
\int_{p_T}^{p^+}\frac{d{\bar\mu}}{\bar\mu}\right]
\nonumber \\
&=&\exp\left[-{\bar\alpha_s}
\int_{(zq)^2}^{Q^2}\frac{dp^2}{p^2}
\int_{0}^{1-p_T/p}\frac{dz'}{1-z'}\right]
\nonumber \\
\Delta_{NS}(z,q,p_T)&=&\exp\left[-2{\bar\alpha_s}
\int_{p_T}^{zq}\frac{dp^+}{p^+}
\int_{p_T}^{p^+}\frac{d{\bar\mu}}{\bar\mu}\right]\;.
\nonumber \\
&=&\exp\left[-{\bar\alpha_s}\int_{z}^{p_T/q}\frac{dz'}{z'}
\int_{(z'q)^2}^{p_T^2}\frac{dp^2}{p^2}\right]\;,
\label{nons}
\end{eqnarray}
where the variable changes ${\bar \mu}=(1-z')p$ and $p^+=p$ for $\Delta_S$, 
and ${\bar \mu}=p$ and $p^+=z'q$ for $\Delta_{NS}$ have been employed. The 
inserted scale $zq$ reflects the combination of the rapidity ordering for 
the BFKL equation and the transverse momentum ordering for the GLAP 
equation \cite{CCFM,KMS}. 


Picking up the last term of the triple-gluon vertex in Eq.~(\ref{tri}),
$\tilde F$ is written, in terms of Fig.~4(b), as
\begin{eqnarray}
{\tilde F}(x,p_T,p^+)&=&-\frac{i}{2}Ng^2\int_x^1 d\xi
\int\frac{d^4l}{(2\pi)^4}
\frac{v^\mu{\hat n}_\mu}{v\cdot l n\cdot l}
2\pi i\delta(l^2)\Delta^2(Q,p_T)
\nonumber \\
& &\times 
\delta(\xi-x-l^+/p^+)\theta(Q-zq)
F(\xi,|{\bf p}_T+{\bf l}_T|,p^+),
\label{ci}
\end{eqnarray}
with $l$ the gluon momentum. 
The propagator $1/v\cdot l$ comes from the eikonalized tree-level 
$J^{(0)}$ on the right-hand side of Fig.~4(b). 
The left-hand side $J^{(0)}$ has been absorbed into 
$F$. The argument of $F$ in the integrand is ${\bf p}_T+{\bf l}_T$, because 
it is for the real gluon contribution. 
Basically, the above formula is similar to
the real gluon part of the BFKL equation (\ref{kf}) except for the 
exponential $\Delta^2$ from the two jet functions $J$, and 
$\delta(\xi-x-l^+/p^+)$ associated with the final state cut, which is 
restored due to the inclusion of the $p^+$ dependence as in Eq.~(\ref{kj}). 
Through this $\delta$ function, the GLAP splitting function
is generated. The $\theta$ function guarantees that the Sudakov exponential 
$\Delta_S$ is meaningful. Hence, the expression of Eq.~(\ref{ci}) 
manifests the combination of the GLAP and BFKL features.

Performing the integration over $l^+$ and $l^-$, we obtain
\begin{eqnarray}
{\tilde F}(x,p_T,p^+)&=&\frac{\bar\alpha_s}{4}\int_x^1 d\xi
\int\frac{d^2l_T}{\pi}
\frac{2n^2(\xi-x)p^{+2}}{[n^+l_T^2+2n^-(\xi-x)^2p^{+2}]^2}\Delta^2(Q,p_T)
\nonumber\\ 
& &\times \theta(Q-zq)F(\xi,|{\bf p}_T+{\bf l}_T|,p^+)\;,
\label{cctf}
\end{eqnarray}
where we have assumed $n=(n^+,n^-,{\bf 0})$ for convenience.
Eq.~(\ref{cctf}) is then substituted into (\ref{cc}) to find the solution
of $F$. We adopt the variable changes $\xi=x/z$ and ${\bf l}_T=
(1-z){\bf q}$, and integrate Eq.~(\ref{cctf}) from $p^+=0$ to 
$Q$. To work out the $p^+$ integration, $F(x/z,|{\bf p}_T+{\bf l}_T|,p^+)$
is approximated by $F(x/z,|{\bf p}_T+{\bf l}_T|,l_T)$. This approximation is 
fine, if one does not intend to obtain the $\delta$-function terms of the 
splitting function $P_{gg}$, written as
\begin{equation}
P_{gg}={\bar\alpha_s}
\left[\frac{1}{1-z}+\frac{1}{z}+z(1-z)+\left(\frac{11}{12}N
-\frac{1}{6}n_f\right)\delta(1-z)\right]\;,
\label{pgg}
\end{equation}
$n_f$ being the number of quark flavors. Eq.~(\ref{cctf}) then becomes 
\begin{eqnarray}
F(x,p_T,Q)&=&F^{(0)}+{\bar\alpha_s}\int_x^1 dz
\int\frac{d^2q}{\pi q^2}\theta(Q-zq)
\Delta_S(Q,zq)\Delta_{NS}(z,q,p_T)
\nonumber \\
& &\hspace{2.0cm}\times
\frac{1}{z(1-z)}F(x/z,|{\bf p}_T+(1-z){\bf q}|,l_T)\;,
\label{ccc1}
\end{eqnarray}
where the nonperturbative initial condition $F^{(0)}$ corresponds to the 
lower bound of $p^+$, and the term suppressed by $1/Q^2$ in the integral 
has been dropped. 

Eq.~(\ref{ccc1}) can be rewritten as
\begin{eqnarray}
F(x,p_T,Q)&=&F^{(0)}+\int_x^1 dz
\int\frac{d^2q}{\pi q^2}\theta(Q-zq)
\Delta_S(Q,zq){\tilde P}(z,q,p_T)
\nonumber \\ 
& &\hspace{2.0cm}\times
F(x/z,|{\bf p}_T+(1-z){\bf q}|,l_T)\;,
\label{ccfm}
\end{eqnarray}
with the splitting function 
\begin{equation}
{\tilde P}={\bar\alpha_s}
\left[\frac{1}{1-z}+\Delta_{NS}(z,q,p_T)\frac{1}{z}+z(1-z)\right]\;.
\end{equation}
To arrive at the above expression we have employed the identity $1/(z(1-z))
\equiv 1/(1-z) +1/z$, and neglected the non-Sudakov form factor 
$\Delta_{NS}$ in front of $1/(1-z)$, because $\Delta_{NS}$ has no pole as 
$z\to 1$ as shown in Eq.~(\ref{nons}). Obviously, Eq.~(\ref{ccfm}) is the 
CCFM equation \cite{CCFM}. The last term $z(1-z)$ of $\tilde P$ is put in 
by hand, which is hinted by Eq.~(\ref{pgg}). This term, finite at $z\to 0$ 
and at $z\to 1$, can not be obtained in the conventional approach 
either \cite{CCFM} as stated at the end of Sec. II.

\vskip 1.0cm

\centerline{\large\bf V. A MODIFIED BFKL EQUATION}
\vskip 0.5cm

We have mentioned in the Introduction that the recent HERA data of the DIS 
structure function $F_2(x,Q^2)$ can be explained by the CCFM equation 
\cite{KMS}, in which the Sudakov exponential $\Delta_S$ collects the 
summation of $\ln Q$ for a large $x$. In this section we propose a modified 
BFKL equation based on the resummation technique. This modified equation, 
much simpler than the CCFM equation, contains an intrinsic $Q$ dependence, 
instead of that from the $\ln Q$ summation. It will be shown that the 
resultant predictions for $F_2$ match the HERA data. With this success, we 
demonstrate the power of the resummation technique. 

An alert reader may have noticed that to derive the BFKL equation, we must 
extend the loop momentum $\l^+$ to infinity in Eq.~(\ref{kf}). Strickly
speaking, the real gluon emission in fact involves the distribution function
$F(x+l^+/p^+,|{\bf p}_T+{\bf l}_T|)$ as part of the integrand, which is then 
approximated by $F(x,|{\bf p}_T+{\bf l}_T|)$ in the soft $l$ region. 
Therefore, the behavior of $F$, vanishing at the momentum fraction equal to
unity, should introduce an upper bound of $l^+$. To obtain a more reasonable 
BFKL kernel, we truncate $l^+$ at some scale, and a plausible choice 
of this scale is of order $Q$. We then derive a modified BFKL equation
for a $Q$ dependent gluon distribution function,
\begin{eqnarray}
\frac{dF(x,p_T,Q)}{d\ln(1/x)}
&=&{\bar\alpha}_s\int\frac{d^{2}l_T}{\pi l_T^2}
[F(x,|{\bf p}_T+{\bf l}_T|,Q)-\theta(Q_0^2-l_T^2)F(x,p_T,Q)]
\nonumber\\
& &-{\bar\alpha}_s\int\frac{d^{2}l_T}{\pi}
\frac{F(x,|{\bf p}_T+{\bf l}_T|,Q)}{l_T^2+Q^{2}}\;,
\label{mbf}
\end{eqnarray}
in the $x\to 0$ limit, where the last term comes from the upper bound of 
$l^+$. The gauge vector $n$ has been chosen to render the coefficient of 
$Q^2$ equal to unity. It can be shown that our predictions are insensitive 
to this coefficient, as long as it is of order unity. 
Obviously, Eq.~(\ref{mbf}) approaches Eq.~(\ref{bfkl}) in the 
$Q\to \infty$ limit. 

Another modification 
is that the loop momentum $l_T$ in the virtual gluon emission is 
truncated at an arbitrary scale $Q_0$ of order 1 GeV, instead of $p_T$ as in 
the conventional BFKL equation. This modification is reasonable, since the 
virtual gluon contribution plays the role of a soft regulator for the real 
gluon emission only, and setting the cutoff to $Q_0$ serves the same 
purpose. Furthermore, the replacement of $p_T$ by $Q_0$ allows us to solve 
Eq.~(\ref{mbf}) analytically, and the solution of $F$ maintains 
the essential BFKL features. 
It will be shown that the first term of the integral 
is responsible for the rise of $F$, and the 
last term acts to slower the rise. We then expect that $F$ ascends faster at 
a larger $Q$, for which the effect of the last term is weaker.

We observe that the CCFM equation will be identical to 
the modified BFKL equation, if the exponential $\Delta^2$
is ``unfolded" into the lowest-order virtual gluon contribution, 
$\theta(Q-zq)$ is dropped, $F(\xi,|{\bf p}_T+{\bf l}_T|,p^+)$ is replaced by 
$F(x,|{\bf p}_T+{\bf l}_T|,Q)$, and the $\xi$ integration is performed to
produce the $Q$ dependent term. Therefore, the $Q$ dependence of the latter
is not attributed to the all-order $\ln Q$ summation in
$\Delta$, but to the boundary of the phase space for radiative corrections.
In fact, this $Q$ dependent term also appears in the CCFM equation, but was  
dropped in the derivation of Eq.~(\ref{ccc1}).

The Fourier transform of Eq.~(\ref{mbf}) leads to
\begin{eqnarray}
\frac{d{\tilde F}(x,b,Q)}{d\ln(1/x)}
&=&{\bar\alpha}_s(1/b)\int\frac{d^{2}l_T}{\pi}
\left[\frac{e^{-i{\bf l}_T\cdot {\bf b}}-\theta(Q_0^2-l_T^2)}{l_T^2}
{\tilde F}(x,b,Q)\right.
\nonumber\\
& &\left.-\frac{e^{-i{\bf l}_T\cdot {\bf b}}}{l_T^2+Q^{2}}
{\tilde F}(x,b,Q)\right]\;,
\nonumber\\
&=&-S(b,Q){\tilde F}(x,b,Q)\;,
\label{bfb}
\end{eqnarray}
with 
\begin{equation}
S(b,Q)=2{\bar\alpha}_s(1/b)[\ln(Q_0 b)+\gamma-\ln 2+K_0(Qb)]\;.
\label{es}
\end{equation}
The Bessel function $K_0$ comes from the last term of the integral, and 
$\gamma$ is the Euler constant. The argument of $\alpha_s$ has 
been chosen as $1/b$, since we work in the conjugate $b$ space. 

Eq.~(\ref{bfb}) can be trivially solved to give
\begin{eqnarray}
{\tilde F}(x,b,Q)={\tilde F}(x_0,b)\exp[-S(b,Q)\ln(x_0/x)]\;,
\label{sbfb}
\end{eqnarray}
$x_0$ being the initial momentum fraction below which 
$F$ begins to evolve according to the BFKL summation.
Transforming Eq.~(\ref{sbfb}) back to the momentum space, we derive
the analytical solution to the modified BFKL equation,
\begin{eqnarray}
F(x,p_T,Q)&=&\int_0^\infty bdb J_0(p_Tb){\tilde F}(x,b,Q)\;,
\label{sbf}
\end{eqnarray}
and the gluon density $xg$ by integrating Eq.~(\ref{sbf}) over $p_T$,
\begin{eqnarray}
xg(x,Q^2)=\int\frac{d^2 p_T}{\pi}F(x,p_T,Q)=
2Q\int_0^{\infty} db J_1(Qb){\tilde F}(x,b,Q)\;.
\label{gbf}
\end{eqnarray}
In the above expressions $J_0$ and $J_1$ are the zeroth and first order
Bessel functions, respectively.

To proceed with the numerical analysis, we assume a ``flat" gluon 
distribution function \cite{KMS,CG}
\begin{equation}
{\tilde F}^{(0)}(x,b)=3N_g(1-x)^5\exp(-Q_0^2b^2/4)\;,
\end{equation}
for $x \ge x_0$, $N_g$ being a normalization constant, which is the Fourier 
transform of 
\begin{equation}
F^{(0)}(x,p_T)=\frac{6}{Q_0^2}N_g(1-x)^5\exp(-p_T^2/Q_0^2)
\end{equation}
in momentum space. The initial condition ${\tilde F}(x_0,b)$ in 
Eq.(\ref{sbfb}) is then equated to 
${\tilde F}(x_0,b)={\tilde F}^{(0)}(x_0,b)$. We set $Q_0=1$ 
GeV and $x_0=0.1$ arbitrarily. It can be shown that the predictions vary 
only slightly for other choices of the parameters $Q_0$ and $x_0$ of the 
same order. $N_g$ will be determined by the data of the structure function 
$F_2(x,Q^2)$ at a specific value of $Q^2$, and then employed to make 
predictions for other values of $Q^2$.

An advantage of the $b$ space is that the infrared sensitivity of the BFKL 
solution from the $p_T$ diffusion is moderated. The divergence of 
$\alpha_s(1/b)$ from a large $b$ is 
suppressed by the exponential $e^{-Q_0^2b^2/4}$ in the initial condition 
${\tilde F}^{(0)}$. In the momentum space, however, the divergence of 
$\alpha_s(p_T)$ from small $p_T$ is not suppressed by $\exp(-p_T^2/Q_0^2)$ 
in $F^{(0)}$, and thus the solution is sensitive to the infrared cutoff
of $p_T$. We have confirmed that our predictions of $F_2$ almost remain
the same for the cutoff of $b$ at 2-4 GeV$^{-1}$.

We compute the structure function $F_2$, whose expression, according to 
the $p_T$-factorization theorem, is given by
\begin{equation}
F_2(x,Q^2)=\int_x^1 \frac{d\xi}{\xi}\int_0^{p_c} \frac{d^2p_T}{\pi}
H(x/\xi,p_T,Q)F(\xi,p_T,Q)\;,
\label{f2}
\end{equation}
$p_c$ being the upper bound of $p_T$ which will be specified later.
The hard scattering subamplitude $H$ denotes the contribution from the quark
box diagrams, where both the incoming photon and gluon are off shell with 
$q^2=-Q^2$ and $p^2=-p_T^2$, respectively. For simplicity, we 
consider only the contraction $-g^{\mu\nu}W_{\mu\nu}$ in the calculation of 
$H$, and neglect the contribution from $p^\mu p^\nu W_{\mu\nu}$, 
which is less important, $W_{\mu\nu}$ being the DIS hadronic tensor. 
We also assume that a charm quark is massless, and that a $b{\bar b}$ quark 
pair is not involved in the box diagram, {\it i.e.} the active flavor number 
$n_f$ in the running coupling constant $\alpha_s$ is equal to 4. Following 
these assumptions, we concentrate on the range of $Q^2$ between 8 and 20
GeV$^2$.

A simple calculation gives 
\begin{eqnarray}
H(z,p_T,Q)&=&e_q^2\frac{\alpha_s}{2\pi}z
\Biggl\{\left[z^2+(1-z)^2-2z(1-2z)\frac{p_T^2}{Q^2}
+2z^2\frac{p_T^4}{Q^4}\right]
\nonumber \\
& &\times
\frac{1}{\sqrt{1-4z^2p_T^2/Q^2}}
\ln\frac{1+\sqrt{1-4z^2p_T^2/Q^2}}{1-\sqrt{1-4z^2p_T^2/Q^2}}-2\Biggr\}\;.
\end{eqnarray}
with $e_q$ the electric charge of the quark $q$.
Note that the terms in the braces approach the splitting function 
\begin{equation}
P_{qg}(z)=\frac{1}{2}[z^2+(1-z)^2] 
\end{equation}
in the $p_T\to 0$ limit. To require a meaningful
$H$, we modifiy the upper bound of $p_T$ in Eq.~(\ref{f2}) from $p_c=Q$ to 
\begin{equation}
p_c=\min\left(Q,\frac{\xi}{2x}Q\right)\;.
\end{equation}

We evaluate the integral in Eq.~(\ref{f2}) straightforwardly for $Q^2=15$
GeV$^2$, and then determine the normalization constant 
$N_g=3.656$ from the data fitting. When $Q^2$ 
varies, we adjust $N_g$ such that $xg$ has a fixed normalization
$\int_0^1 xgdx$. It is found that $N_g$ changes only by about 5\% in the 
considered range of $Q^2$. $F_2$ for $Q^2=8.5$, 12, and 20 GeV$^2$ are 
then computed, and results along with the HERA data \cite{H1} are displayed 
in Fig.~5. It is obvious that our predictions agree with the data well. The 
curve has a steeper rise at a larger $Q$, which is the consequence of the 
$Q$ dependent modified BFKL equation. The curves descend rapidly at $x$ 
close to 0.1, since the contributions from other kinds of partons, such as 
the valence quarks, which are more important in this intermediate $x$ region, 
are not included. For comparision, we also present the results from the 
conventional BFKL equation, which can be obtained simply by substituting 
$l_T^2+M^2$ for the denominator $l_T^2+Q^2$ in Eq.~(\ref{mbf}), or 
equivalently, $Mb$ for the argument $Qb$ of the Bessel function $K_0$ in 
Eq.~(\ref{es}) with an extremely large $M=10^3$-$10^4$ GeV. In this case the 
normalization constant determined from the best fit to the data for $Q^2=15$ 
GeV$^2$ is $N_g=2.908$. It is found that the shape of the curves is 
almost independent of $Q$, and thus the match with the data is not very
satisfactory.

At last, we show in Fig.~7 the behavior of the gluon density $xg$ computed 
from Eq.~(\ref{gbf}). The rise of $xg$ at small $x$ is due to the term 
$\ln Q_0b$ in the exponent $S$ as mentioned before, which leads to an 
integrand proportional to 
\begin{equation}
(Q_0b)^{-2\alpha_s\ln(x_0/x)}\;. 
\end{equation}
The small $b$ region then 
gives a huge contribution. When $x$ approaches zero such that 
$2\alpha_s\ln(x_0/x)>1$, the integration over $b$ diverges. However, the 
last term of $S$, $K_0(Qb)\propto -\ln(Qb)$ in the $b\to 0$ limit, which 
comes from the upper bound of the loop momentum $l^+$, cancels the 
divergence. This is the reason $l^+$ must be truncated. On the other hand, 
the variation of $xg$ with $x$ for different $Q$ can also be understood from
the combination of $\ln(Q_0 b)$ and $K_0(Qb)$ in the $b\to 0$ limit, 
written as,
\begin{equation}
\left(\frac{x}{x_0}\right)^{-2\alpha_s\ln(Q/Q_0)}\;.
\end{equation}
The above expression indicates that the exponent $\lambda$,
characterizing the rise of $xg\sim x^{-\lambda}$ at small $x$, increases 
with $Q$. The values $\lambda\approx 0.36$ for $Q^2=8.5$ GeV$^2$ and 
$\lambda\approx 0.51$ for $Q^2=20$ GeV$^2$ are deduced from Fig.~6, which 
are consistent with that obtained in \cite{MSR} from a phenomenological
fit to the HERA data ($\lambda\approx 0.3$ for $Q^2=4$ GeV$^2$) and 
with that in \cite{AKMS} from solving the conventional
BFKL equation numerically ($\lambda\approx 0.5$ for a wide range of $Q^2$).

\vskip 1.0cm

\centerline{\large\bf VI. CONCLUSION}
\vskip 0.5cm

In this paper we have shown that the resummation technique provides a 
unified and simple viewpoint to the organization of the various large
logarithms, and reduces to the GLAP equation, the BFKL equation, and the 
CCFM equation in different kinematic regions. The main idea is to relate the 
derivative of a parton distribution function to a new function involving a 
special vertex. The summation of the large logarithms is then embedded in 
the new function without resort to the complicated diagrammatic analyses. 
When expressing the new function as a factorization formula, we obtain the 
evolution equation, and the subdiagram containing the special vertex is 
exactly the corresponding kernel. By means of the resummation technique, the 
derivation of the evolution equations is simpler. Furthermore, to improve 
the accuracy of the kernel to next-to-leading logarithms, we only need to 
evaluate the $O(\alpha_s^2)$ subdiagram. Such an evaluation can be performed 
in a straightforward way. The BFKL equation including the summation of the
next-to-leading $\ln(1/x)$ will be published elsewhere.

We have also calculated the DIS structure function $F_2(x,Q^2)$ 
using the analytical solution of the 
unintegrated gluon distribution function from the modified BFKL equation. 
Our predictions exhibit a stronger $Q$ dependence, and are in a good
agreement with the HERA data, compared to those from the conventional
BFKL equation. Note that the $Q$ dependence in the modified BFKL equation
is intrinsic, which arises from the boundary of the phase space for
radiative corrections, instead of from the $\ln Q$ summation in the CCFM 
equation. In this sense we argue that the current experiemnts 
may have explored the multi-Regge region with $\ln(1/x) \gg \ln Q$.
Certainly, this issue still needs to be clarified by further theoretical 
and experimental studies \cite{KLM}.

\vskip 0.5cm
This work is supported by National Science Council of R.O.C. under the 
Grant No. NSC-86-2112-M-194-007.

\newpage
\centerline{\large \bf Figure Captions}
\vskip 0.5cm

\noindent
{\bf FIG. 1.} Definition of a parton distribution function in (a) the
covariant gauge and in (b) the axial gauge.
\vskip 0.5cm

\noindent
{\bf FIG. 2.} (a) The derivative $p^+d\phi/dp^+$ in the axial gauge.
(b) The $O(\alpha_s)$ function $K$. 
(c) The $O(\alpha_s)$ function $G$. 
\vskip 0.5cm

\noindent
{\bf FIG. 3.} (a) The derivative $-xdF/dx$ in the covariant gauge.
(b) The soft structure and (c) the ultraviolet structure of the
$O(\alpha_s)$ subdiagram containing the special vertex.
\vskip 0.5cm

\noindent
{\bf FIG. 4.} (a) The subdiagram containing the special vertex for the CCFM 
equation. (b) The subdiagram for the CCFM equation after resumming the 
double logarithms in $J$. 

\vskip 0.5cm

\noindent
{\bf FIG. 5.} The dependence of $F_2$ on $x$ derived 
from the modified BFKL equation (solid lines) and from the conventional
BFKL equation (dashed lines). The HERA data \cite{H1} are also shown.
\vskip 0.5cm

\noindent
{\bf FIG. 6.} The dependence of $xg$ on $x$ derived from
the modified BFKL equation for, from bottom to top, $Q^2=8.5$, 12, 15,
and 20 GeV$^2$.

\end{document}